\documentclass[sigconf]{acmart}
\copyrightyear{2024}
\acmYear{2024}
\setcopyright{rightsretained}
\acmConference[AM '24]{Audio Mostly 2024 - Explorations in Sonic Cultures}{September 18--20, 2024}{Milan, Italy}
\acmBooktitle{Audio Mostly 2024 - Explorations in Sonic Cultures (AM '24), September 18--20, 2024, Milan, Italy}
\acmPrice{}
\acmDOI{10.1145/3678299.3678321}
\acmISBN{979-8-4007-0968-5/24/09}

\begin{document}

\title{Tonal Cognition in Sonification: Exploring the Needs of Practitioners in Sonic Interaction Design}

\author{Minsik Choi}
\email{minsik.choi@anu.edu.au}
\authornotemark[1]
\affiliation{%
  \institution{School of Computing, The Australian National University}
  \city{Canberra}
  \state{ACT}
  \country{Australia}
  \postcode{2601}
}

\author{Josh Andres}
\email{josh.andres@anu.edu.au}
\authornotemark[2]
\affiliation{%
  \institution{School of Cybernetics, The Australian National University}
  \city{Canberra}
  \state{ACT}
  \country{Australia}
  \postcode{2601}
}

\author{Charles Patrick Martin}
\email{charles.martin@anu.edu.au}
\authornotemark[1]
\affiliation{%
  \institution{School of Computing, The Australian National University}
  \city{Canberra}
  \state{ACT}
  \country{Australia}
  \postcode{2601}
}

\renewcommand{\shortauthors}{Choi, et al.}

\begin{abstract}

Research into tonal music examines the structural relationships among sounds and how they align with our auditory perception. The exploration of integrating tonal cognition into sonic interaction design, particularly for practitioners lacking extensive musical knowledge, and developing an accessible software tool, remains limited. We report on a study of designers to understand the sound creation practices of industry experts and explore how infusing tonal music principles into a sound design tool can better support their craft and enhance the sonic experiences they create. Our study collected qualitative data through semi-structured individual and focus group interviews with six participants. We developed a low-fidelity prototype sound design tool that involves practical methods of functional harmony and interaction design discussed in focus groups. We identified four themes through reflexive thematic analysis: decision-making, domain knowledge and terminology, collaboration, and contexts in sound creation. Finally, we discussed design considerations for an accessible sonic interaction design tool that aligns auditory experience more closely with tonal cognition.

\end{abstract}

\ccsdesc[500]{Human-centered computing~User studies}
\ccsdesc[500]{Human-centered computing~Systems and tools for interaction design}

\keywords{sonic interaction design, auditory display, creativity support tools, user-centered design, embodied music cognition, tonal cognition}

\begin{teaserfigure}
   \includegraphics[width=\textwidth]{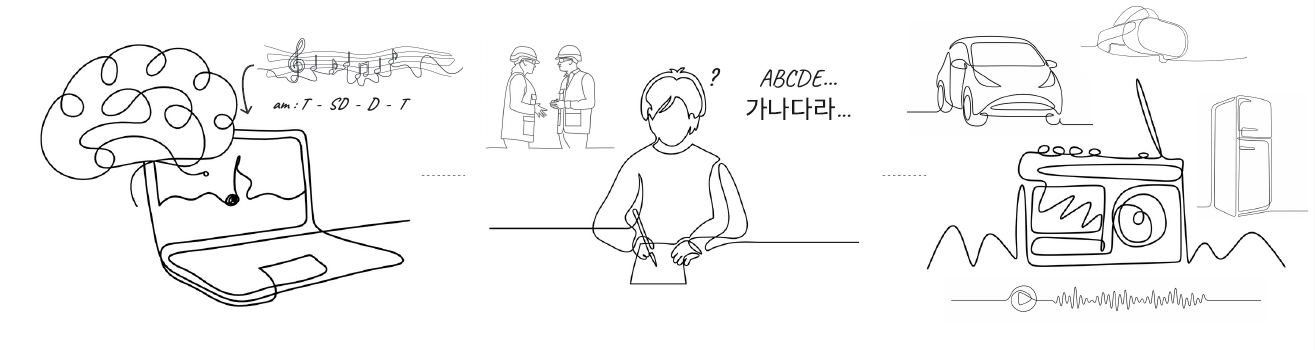}
   \caption{How can tonal music principles and the needs of practitioners be incorporated in sonic interaction design tools?}
\end{teaserfigure}

\maketitle

\section{Introduction}

The study of tonal music explores the underlying structural relationships among sounds~\cite{Grout2019}. It has evolved into a systematic framework for hierarchically organizing individual sounds within a continuous perspective~\cite{christensen2006cambridge}. Insights from cognitive science, such as musical expectancy and affordance, which evoke sonic tension and release and impact subsequent actions, have further emphasized the cognitive flow that listeners experience from tonal music as structured sound~\cite{koelsch2012brain, bigand2016tonal, korsakova2018two}. For instance, akin to how humans grasp the harmonic relationships in simple melodies, applying tonal cognition offers listener-centered avenues for crafting auditory displays of everyday products~\cite{roddy2020mapping, serafin2022auditory,boehringer2022listener}.

Interest in tonal cognition arose along with music cognitive applications in interactive sonification~\cite{hermann2011sonification, leman2008embodied}. Gaver~\cite{gaver1, gaver2} introduced the concept of auditory icon, drawing upon everyday auditory cognition, and Norman~\cite{norman2013design} briefly explored sound signals as an alternative to visual displays in a similar context. Blattner et al.~\cite{Blattner1989} categorized communicative sounds known as earcons by musical elements like rhythm and pitch, and Brewster explored in depth across studies such as~\cite{brewster1994detailed, brewster1995experimentally, brewster1998using}. Furthermore, Serafin et al.~\cite{serafin2022auditory} expanded the discourse by highlighting the significance of auditory interaction within musical structures~\cite{franinovic2013sonic}, and Vickers~\cite{vickers2016sonification} addressed the intricate relationship between music and sonification~\cite{scaletti2018sonification}, mentioning tonal functions and embodied cognition. Research on its application accordingly explored movement sonification between embodied music cognition and task sequence such as Newbold et al.~\cite{newbold2016, newbold2017musical}. These confirm user's auditory experience based on musical principles, leading to the discussion on tonal cognition and its employment in interactive sonification. 

The application of tonal cognition to sonic interaction, however, remains limited~\cite{choi2023music}. There is a scarcity of auditory interface design research for designers lacking extensive musical knowledge, and few instances of software solutions aimed at enhancing accessibility of tonal music principles. Interdisciplinary practitioners in auditory interface design process can further complicate the context. Research in interactive sonification has typically focused on end-users for design planning and evaluation~\cite{moesgaard2020involving, boehringer2022listener}, and particularly within the applications of tonal cognition, there has been a lack of pragmatic solutions proposed, along with modest attention to methods in use and accessible platforms for practitioners. Although Newbold et al.~\cite{newbold2020movement} presented a design model for movement altering sonification, hands-on methods of tonal music and interaction design can be incorporated through a software solution to enhance accessibility for practitioners. Better understanding of designers and their practices could facilitate the development of software that assists them in harnessing tonal cognition through sensible methods for interactive sonification.

This user research, therefore, aims to study sound design practitioners to understand their work practices, language usage, and functional needs to build a design tool deploying tonal cognition. Our method involved individual and focus group interviews followed by reflexive thematic analysis and task flow analysis. We produced a low-fidelity prototype of a software tool that would allow designers to access of tonal principles with interaction design—harmonic functions and user tasks—discussed by the focus groups. We identified four themes from overall qualitative data that foreground decision-making, domain knowledge and terminology, collaboration, and design contexts in sound creation. We discuss further designer-centered software solution development directions based on the results. This research is intended with further stages involving creating high-fidelity prototypes and conducting usability evaluations. Our contributions involve unpacking current sound design practices and potential needs for a sonic interaction tool with tonal cognition, and accordingly providing recommendations for the future development of an accessible tool for practitioners in sonic interaction design.

\section{Background}\label{se:background}

Functionality in tonal music frames chordal relationships through three pillars, tonic, dominant and subdominant, where the behavior of chords is in hierarchical relation to the tonic~\cite{Bernstein_2002}. Tonal cognition within these hierarchical relations enables humans to perceive and appreciate tension and release as musical expectancy~\cite{koelsch2012brain}. Korsakova-Kreyn~\cite{korsakova2018two} proposed two distinct levels of embodied cognition in music: corporeal articulation at a surface level and tonal cognition at a deep level, and Vickers~\cite{vickers2016sonification} discussed structured tonal musical frameworks in relation to embodied music cognition and sonification. Newbold et al.~\cite{newbold2020movement} ultimately suggested a design method using concepts of tonal cognition such as musical expectancy for movement sonification and explored this concept through user studies~\cite{newbold2016, newbold2017musical}.

Meanwhile, user-centered design in software development~\cite{medhi2007user} has been applied across domains, including interactive sonification~\cite{hermann2011sonification, wallach2012usercentered}. 
End-users have been the primary focus of user studies, including experimental testing for usability attributes~\cite{sagar2017systematic}, while studies specifically targeting designers have been relatively rare. Some studies have examined designers' perspectives assisting practitioners in design practices~\cite{Daalhuizen2009, burnell2017integrating}. Selfridge and Pauletto~\cite{selfridge2022sound} compared initial design ideas among sound designers to find out common points, and Kameth et al.~\cite{Kamath2024} studied designers using sound design support tools and discussed human-AI design considerations; however, few or no studies have specifically explored designer-centered approaches in sonification methods related to tonal cognition. 
The interdisciplinary nature of sonification in interaction design~\cite{jeon2020auditory, rogers2011interaction} suggests that designer-centered research addressing this gap might have a wide impact on a diverse range of practitioners.

\begin{figure*}
  \centering
  \includegraphics[width=\textwidth]{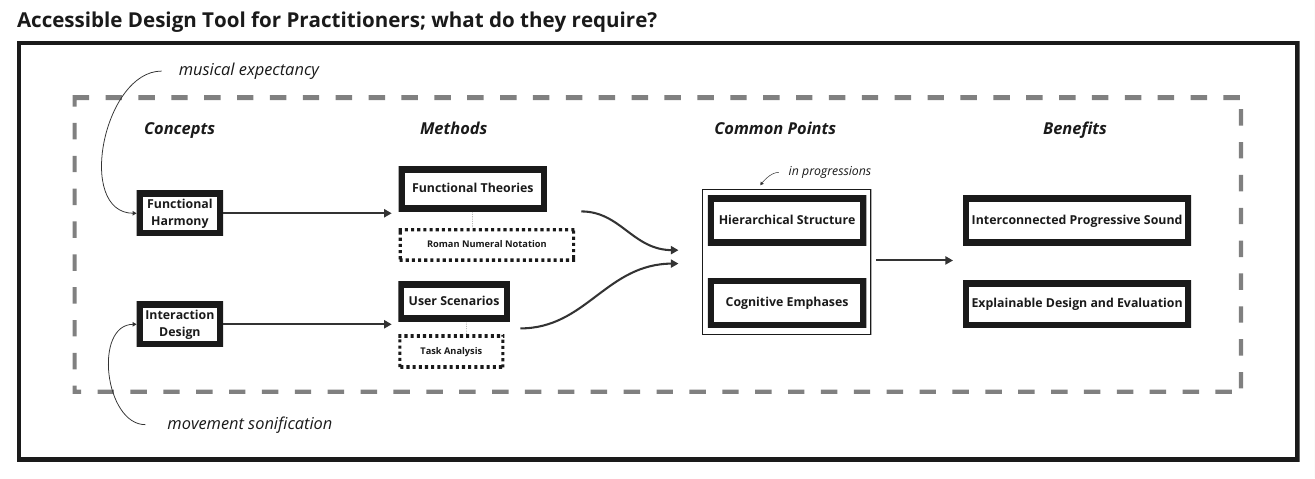}
  \caption{Background overview including fundamental concepts, corresponding methods, common ground, and following benefits; musical expectancy and movement sonification are the main points from Newbold et al.~\cite{newbold2020movement}, and the bigger frame represents the potential software development for designers reflecting this sonification strategy.}
  \label{fg:background}
\end{figure*}

Efforts to refine design theories through practical methods~\cite{lipton2011practical, stanton2017human} imply the potential benefits of hand-on application of tonal cognition in sonification. Tonal cognition is rooted in the hierarchical harmonic functions perceived by audience~\cite{bigand2016tonal}, and interaction design where the concept is applied begins with user understanding with scenarios and subdivided task analysis, resulting in sonic interaction design with tonal cognition~\cite{rogers2011interaction, hermann2011sonification}. The methods in both perspectives share hierarchical structures and emphasize cognitive aspects among individual notes and tasks within continuing contexts~\cite{rogers2011interaction, clement2023rock}. Harmonic functions are codified using Roman numeral notation~\cite{gauldin2004harmonic}, and user scenarios with task analysis enable practitioners to assign appropriate subtasks for sonic interface design~\cite{rogers2011interaction}. Aligning these methods in an application could allow designers pragmatic access to tonal cognition concepts in sonic interaction design. 
In this paper, we intend to explore this alignment, described in Figure~\ref{fg:background}, by studying a low-fidelity prototype sound design tool with design practitioners. This new approach has the potential to facilitate interconnected progressive displays and explainable design plans within the context of accessibility~\cite{choi2023music}.

\section{Method}

\begin{figure*}
  \centering
  \includegraphics[width=\textwidth]{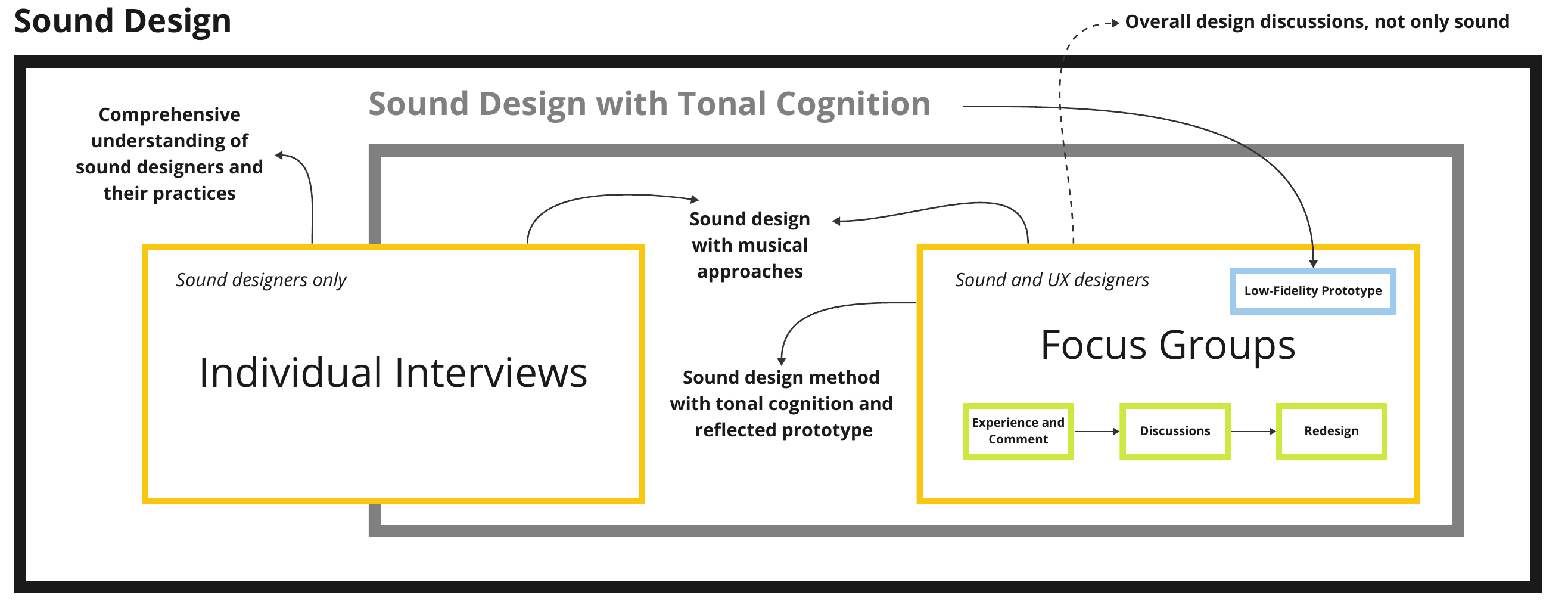}
  \caption{Method overview illustrating conceptual relationships, research scope of each method, and procedure for focus groups. The black frame represents overall sound design, while the gray frame specifically denotes sound design with tonal cognition. The two yellow boxes illustrate what each method encompasses: individual interviews cover both general sound design and sound design with tonal cognition, while focus groups concentrated exclusively on aspects of sound design with tonal cognition using a low-fidelity prototype framed by a blue box. The green boxes outline the procedure of the focus groups.}
  \label{fg:explanation}
\end{figure*}

Our study included semi-structured individual interviews with three sound designers, each lasting 30 minutes, along with one hour-focus group sessions with three sound designers and three UX designers, as outlined in Figure~\ref{fg:explanation}. The individual interviews aimed to gain an extensive understanding of sound designers, encompassing but not limited to musical approaches. Two focus group sessions, individually with sound and UX designers, focused on hands-on experience with a low-fidelity prototype described in Figure~\ref{fg:prototype} and Appendix~\ref{append:figma}. The low-fidelity prototype demonstrated a software tool reflecting the suggested design logic with the methods of functional harmony and interaction design elaborated in section~\ref{se:background}. Both individual and focus group interviews were conducted using predetermined open questions (see Appendix~\ref{appendix:question}) within the recording environment depicted in Figure~\ref{fg:environment}. Sound designers were asked about their work practices, dynamics, and common language usage related to sound creation, including musical approaches, in individual interviews. Focus groups involved written and oral inquiries about the prototype experience, alongside design discussions beyond sonification that embraced broader opinions from UX designers. 

\subsection{Prototype} \label{prototype}

\begin{figure*}
  \centering
  \includegraphics[width=\textwidth]{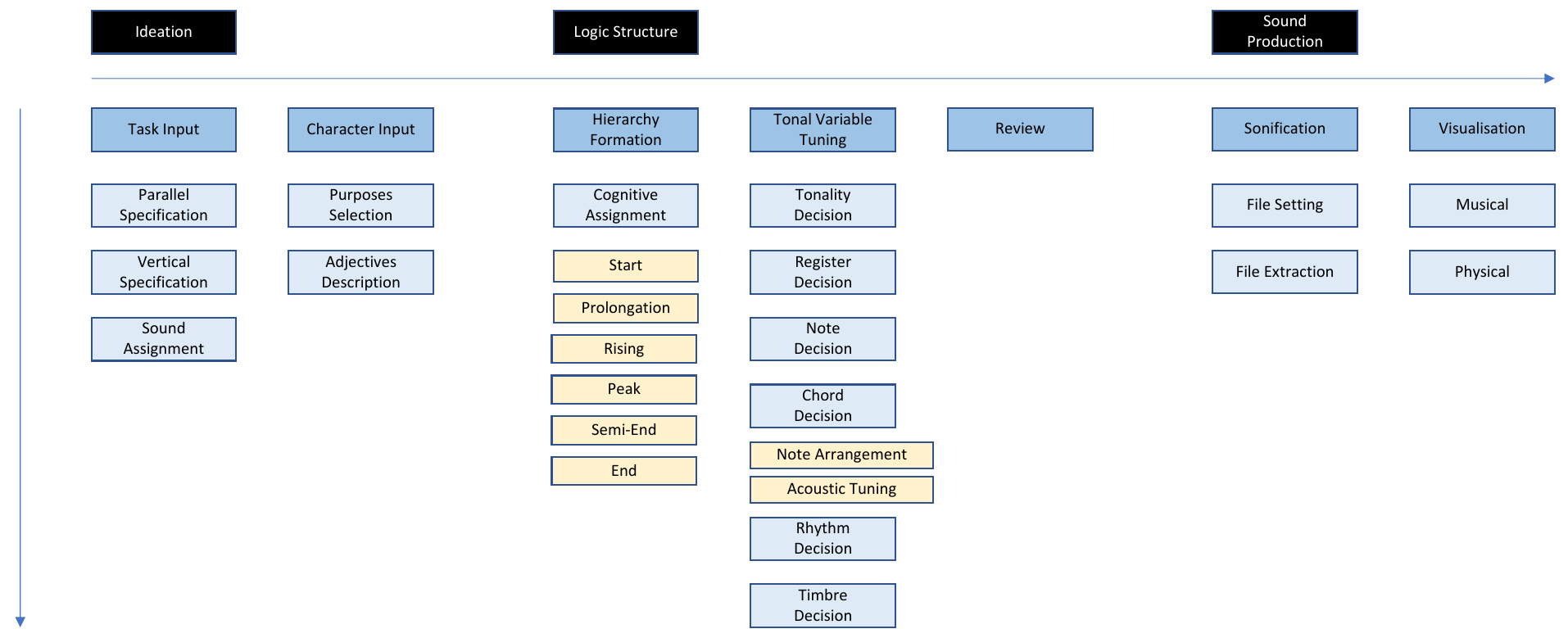}
  \caption{Prototype design logic with user scenario formation, cognitive tonal functions arrangement to target tasks, tonal variable tuning, and file extraction within three main phases.}
  \label{fg:prototype}
\end{figure*}

We developed a low-fidelity prototype of a sonic interaction design tool incorporating tonal cognition with harmonic functions and user tasks. The prototype illustrates the design procedure in three phases: ideation, logic structure, and sound production, with plain language for accessibility, as detailed in Figure~\ref{fg:prototype}. The ideation phase involves creating parallel user scenarios organized in vertical structures, similar to hierarchical task analysis~\cite{stanton2006hierarchical}. Specific purposes and adjectives are assigned to target tasks to guide interface character and effects decisions. The logic structure phase is mainly cognitive arrangement, linking to tonal functions such as start, prolongation, rising, peak, semi-end, and end. Each sound can be tuned by tonal factors for variations in each function. For instance, chord decision includes note arrangement within each combination and acoustic tuning for each note. The sound production phase is for file extraction in audio and visual formats. This phase was designed to allow for open exploration, encouraging sensible ideas. Basic frame was suggested such as file format, sample rate, channel, and bit depth for sonification, and musical and physical—score and spectrogram—are the options for visualization. This prototype was developed using Figma, shown in Appendix~\ref{append:figma}, with a focus on the logical flow and structure of the interface, rather than emphasizing specific visual interface elements.

\subsection{Participants}

To gain a broad understanding of sonic interaction design practices in this initial user research, we recruited three sound designers and three UX designers for our study as described in Table~\ref{tab:participants}.
All six participants were employed at large engineering companies in South Korea. 
Only the three sound designers participated in the individual interviews to explore common sound design contexts, beyond the prototype related to musical approaches. 
Despite belonging to the same company, the sound designers held varied positions with different responsibilities providing diverse insights within the sound design field. The UX designers also contributed overall interface design perspectives as they came from different industries and companies. 
The participants educational backgrounds were diverse yet primarily focused on sound, user experience and related engineering fields. 
Their musical backgrounds also varied but balanced with formal music education, music as ongoing hobby, and brief experiences in musical instruments.

\begin{table*}[]
\caption{Participant information for three UX designers (UX) and three sound designers (SD); HF: human factors, eng.: engineering, and FGI: focus group interview.}
\label{tab:participants}
\resizebox{\textwidth}{!}{%
\begin{tabular}{@{}p{1.5cm}p{2.1cm}p{2.1cm}p{2.1cm}p{2.1cm}p{2.1cm}p{2.1cm}@{}}
\toprule
& {\small P1 (UX)} & {\small P2 (UX)} & {\small P3 (UX)} & {\small P4 (SD)} & {\small P5 (SD)} & {\small P6 (SD)} \\ \midrule
{\small \textbf{Industry}} & {\footnotesize Mobile phone} & {\footnotesize Home appliance} & {\footnotesize Automotive} & {\footnotesize Audio, automotive} & {\footnotesize Audio, automotive} & {\footnotesize Audio, automotive} \\
{\small \textbf{Position}} & {\footnotesize Senior researcher} & {\footnotesize Senior researcher} & {\footnotesize UX researcher} & {\footnotesize Senior manager} & {\footnotesize Senior engineer} & {\footnotesize Technician} \\
{\small \textbf{Role}} & {\footnotesize User research on mobile experience} & {\footnotesize Home appliance UX Planning} & {\footnotesize Vehicle UX design and eng. solution} & {\footnotesize Car audio system layout design} & {\footnotesize Virtual sound and noise tuning} & {\footnotesize Warning sound tuning and EQ validation} \\
{\small \textbf{Education}} & {\footnotesize Psychology, cognitive science, HF} & {\footnotesize Communication design, HF} & {\footnotesize Architecture, HF} & {\footnotesize Electrical eng., composition, HF} & {\footnotesize Electrical eng.} & {\footnotesize Automotive eng.} \\
{\small \textbf{Music Experience}} & {\footnotesize Brief experience in guitar} & {\footnotesize Brief experience in flute} & {\footnotesize Brief experience in violin and bass} & {\footnotesize Majored and choir as ongoing hobby} & {\footnotesize Guitar and recording as ongoing hobby} & {\footnotesize Brief experience in piano} \\
{\small \textbf{Participation}} & {\footnotesize FGI} & {\footnotesize FGI} & {\footnotesize FGI} & {\footnotesize Interview, FGI} & {\footnotesize Interview, FGI} & {\footnotesize Interview, FGI} \\ \bottomrule
\end{tabular}%
}
\end{table*}

\subsection{Procedure}

\begin{figure*}
  \centering
  \includegraphics[width=\textwidth]{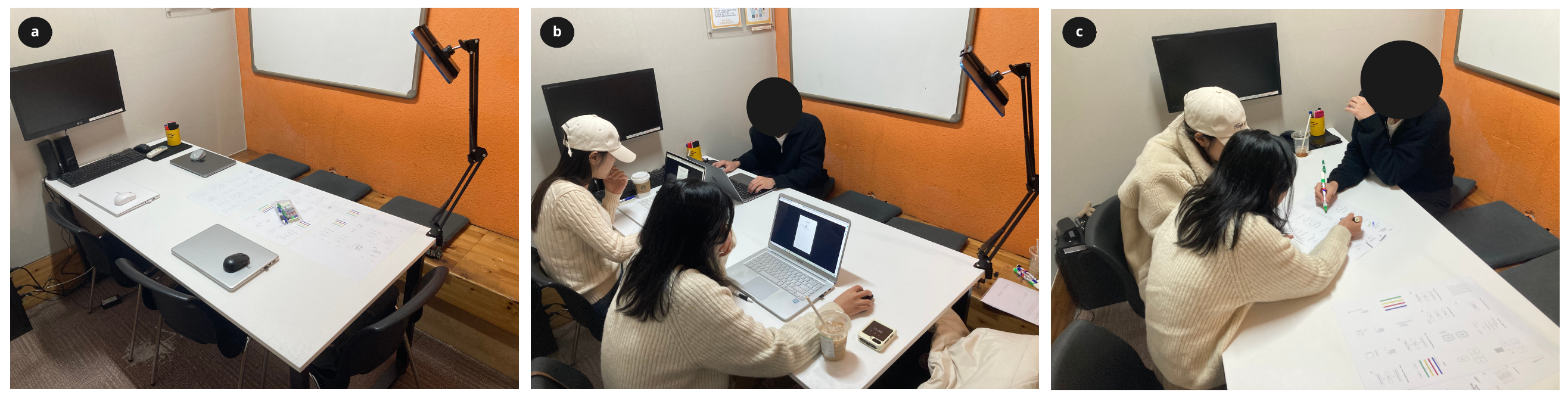}
  \caption{Focus group description; a: interview environment with individual laptops and recording device, b: individual examination of prototype in Figma via Microsoft Forms, c: redesigning activity with paper-based prototype and pens.}
  \label{fg:environment}
\end{figure*}

Individual interviews were conducted in a conversational manner, with questions prompting relevant discussions. The focus group interviews included four steps: exploring the prototype, responding to provided questions via Microsoft Forms with individual laptops based on the prototype, exchanging ideas about the prototype experience through group discussion, and collaboratively redesigning the prototype using paper-based samples and pens captured in Figure~\ref{fg:environment}. They provided feedback on the prototype experience in written, verbal, and visual formats during the focus group sessions. The study commenced with a participant briefing and participants provided written informed consent as well as demographic information. Participants did not receive any incentives for their participation. Ethical approval was obtained from ANON Institutional Review Board.

\subsection{Analysis}\label{analysis-methods}

Qualitative data from individual and focus group interviews were analyzed, resulting in the identification of reflexive themes through thematic analysis~\cite{terry2021essentials}. We employed a bottom-up approach, familiarizing ourselves with the collected qualitative data: interview transcripts, prototype experience responses, and redesign sketches and verbal descriptions. Gathered data was semantically coded with quotes pertaining to design practices and functional needs for the prototype, considering participants' educational and musical backgrounds. This process led us to organize reflexive themes as semantic clusters that overlapped with each other as illustrated in Figure~\ref{fg:reflexive}. We also applied task flow analysis to the work processes of the three sound designers to reveal common tendencies in the way of achieving their goals~\cite{diaper2003handbook}. 
The individual interview data with sound designers was specifically used for task flow analysis, and the outcome is presented in section~\ref{se:task analysis}.

\section{Results} 

\begin{figure*}
  \centering
  \includegraphics[width=\textwidth]{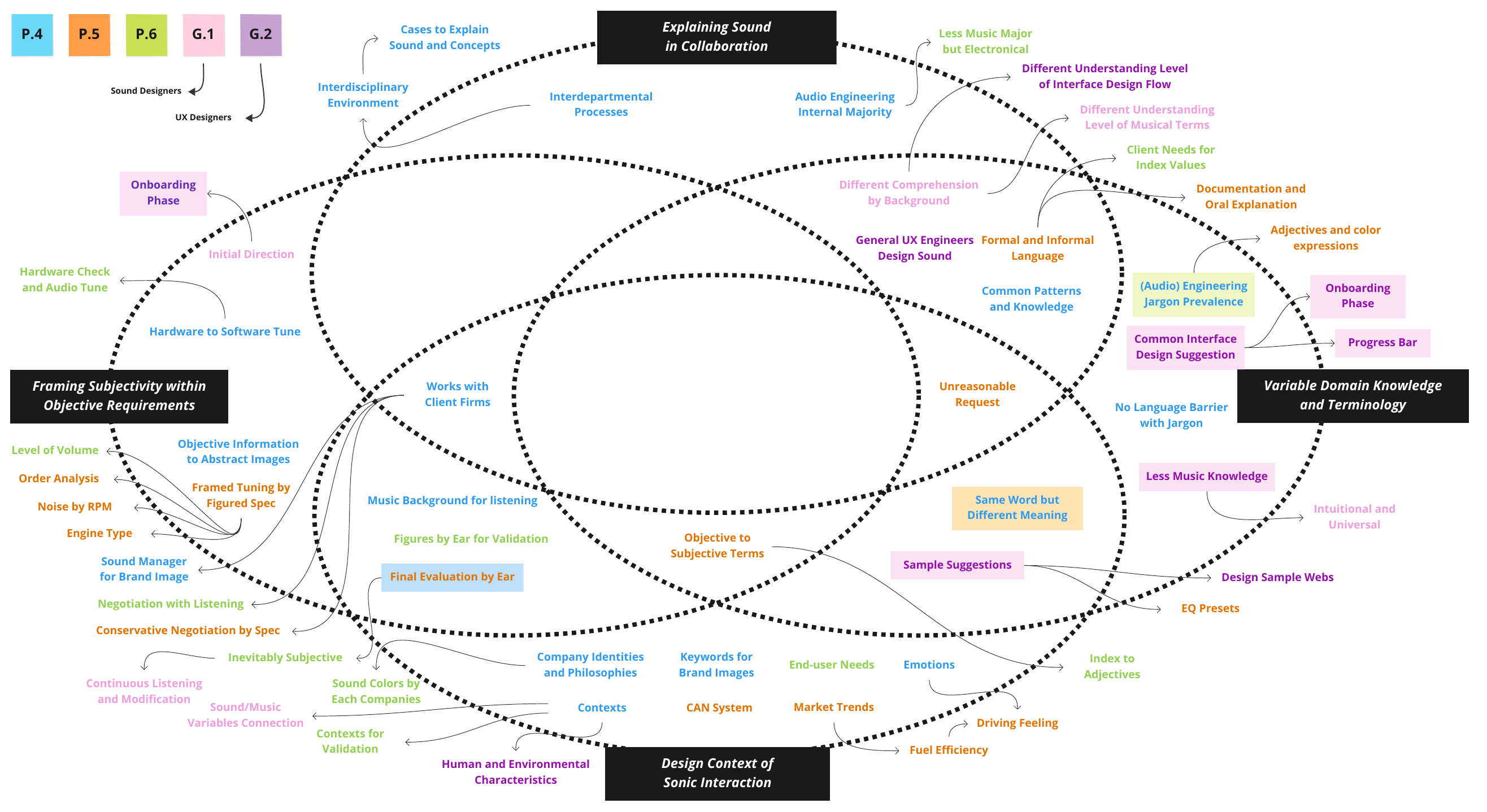}
  \caption{Four discovered reflexive themes comprising individual codes classified by thematic analysis; the colors indicate the source of each code, such as which participant or focus group it originated from, and when a code is mentioned by more than one participant or focus group, two colors are used with additional highlighting.}
  \label{fg:reflexive}
\end{figure*}

In this section, we discuss the themes that were defined by applying our process of reflexive thematic analysis to the qualitative data described in section~\ref{analysis-methods} above.
We defined four themes which are shown as overlapping ovals in Figure~\ref{fg:reflexive}. \textit{Framing subjectivity within objective requirements} concerns the procedural characteristics of sound design practices, moving from concrete criteria to intuitive approaches. 
The sound designers ultimately create sound intuitively based on conceptual keywords, grounded in a concrete and physical foundation. \textit{Variable domain knowledge and terminology} encompasses the language aspects commonly used, primarily domain-specific yet also incorporating universal elements with boundaryless backgrounds and varying levels of music knowledge. \textit{Explaining sound in collaboration} focuses on the need for communicability in interdisciplinary collaboration. Individuals from various backgrounds and musical experience converge for a common purpose, collaborating and eventually reaching a consensus with understandable explanations. \textit{Design context of sonic interaction} describes the contextual aspects involved in designing sound as a whole. These qualitative aspects encompass the entire work process from ideation to generating sound, aiding in the conceptual design direction and sound optimization across diverse situations.

\subsection{Framing Subjectivity within Objective Requirements}\label{se:task analysis}

\begin{figure*}
  \centering
  \includegraphics[width=\textwidth]{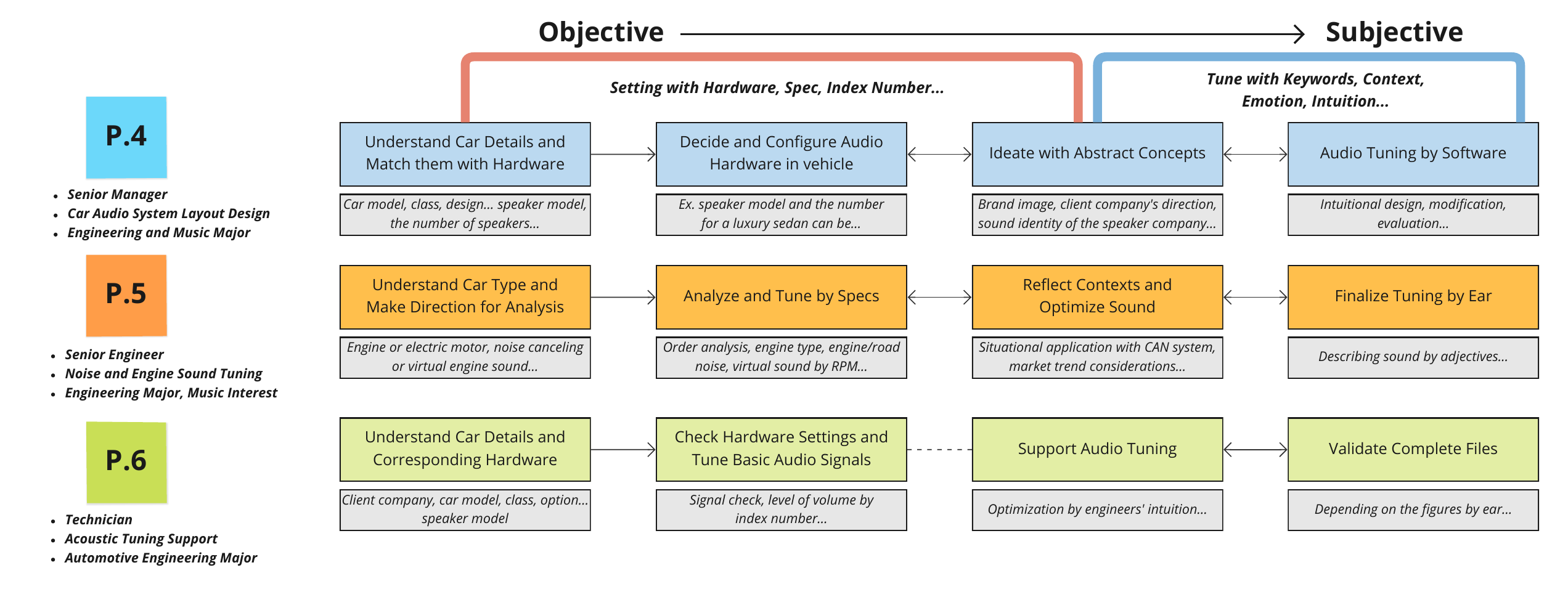}
  \caption{Task flow analysis illustrating the broader framework of designing sound from objectivity to subjectivity; the boxes, color-coded based on each sound designer, show the detailed phases of the sound design process according to their positions and roles.}
  \label{fg:task}
\end{figure*}

Our sound designers begin a project by physical configuration, although the final sound is shaped by subjective intuition. They establish a design foundation through hardware decisions, primarily focusing on speakers, along with numerical specifications, aligning them with target products. After the environmental setting, they proceed to refine the sound based on conceptual imagery, including brand keywords, user contexts and emotions, and their own intuitions. As shown in Figure~\ref{fg:task}, while specific tasks vary depending on roles, their broader framework is characterized by a flow from objective to subjective approaches as the essence of their practices.

P4 described his work process, starting with hardware configuration followed by software tuning. He uses objective information initially and then transitions to more abstract concepts like brand keywords. With a background in music, he emphasized the advantage of his listening skills acquired through music education, particularly for the final tuning stage. P5 primarily engages in framed tuning based on foundational methods, such as order analysis and noise control by RPM. At the final evaluation stage, however, he relies subjectively on his ears and felt experience of the sound. Similarly, P6 begins with basic hardware checks and adjustments as a technician, such as setting the volume level for warning sounds, but he relies on intuition-based figures and abstract concepts by engineers for the broader audio tuning and validation. In both focus groups, the participants recommended incorporating an onboarding phase in the prototype to establish a clear design direction. This aspect is also mentioned by the UX designers and so could be considered universal; however, it is still reasonable to align it with the practices involved in designing sound within this theme.

\subsection{Variable Domain Knowledge and Terminology}

The sound designers utilize jargon specific to the audio engineering field for internal communication. For instance, as P4 highlighted, when communicating with designers from different language backgrounds, there are almost no issues since everyone tends to use similar terminology. In contrast, certain words may hold different connotations within sound design. For example, while loudness may commonly denote the physical strength of sound, it refers specifically to the human-centered perspective of volume. P5 also noted the internal usage of adjectives relating to timbre, such as nasal, as well as color-related analogies for sound such as, bright, dark or neutral gray.

The sound designers, however, basically collaborate, and individual understanding of related domains, such as musical or physical expertise, may vary depending on each background. The UX designers are also occasionally engaged in designing sound, indicating that sound design may not be exclusive to sound designer as a profession. During the focus group with sound designers, there was a discussion about naming music variables, and opinions varied: P4 - "\textit{Tonality could be referred to as harmonics, which is a more understandable engineering term."}, P5 - \textit{"I think both terms are still difficult to understand; perhaps tonality could simply mean a tone?}". While P4, with musical backgrounds, may have a better grasp of tonality and suggested considering the field's subtle nature, P5, who understood fewer musical terms, provided a simpler suggestion that may not correlate with uses within music or sound. Similarly, during the focus group with UX designers, the participants had a stronger understanding of the ideation phase with task flow input in the prototype compared to the sound designers, possibly due to their design-related backgrounds. P2, a home appliance UX designer, shared her experience in auditory interface design, conceptualizing sound with adjectives to guide the development of ring signals. P1 and P3 mentioned parallel experiences in mobile and automotive projects. 

\begin{figure*}
  \centering
  \includegraphics[width=\textwidth]{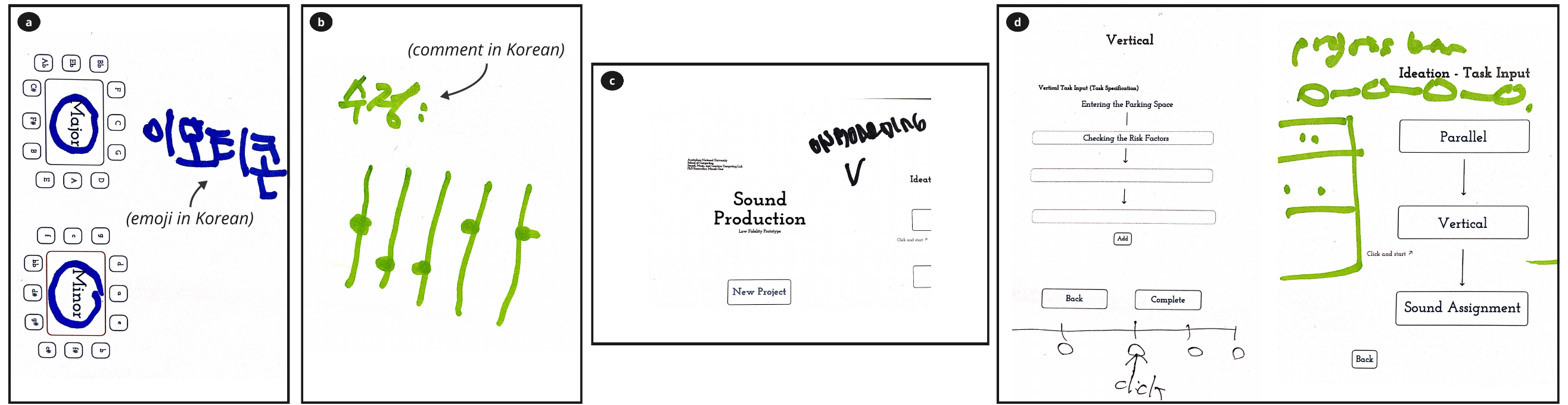}
  \caption{UI element recommendations from the redesigning activity; a: emojis usage for tonality description, b: intuitive control bars, c: onboarding phase, d: progress bars.}
  \label{fg:redesign}
\end{figure*}

The participants had less familiarity with musical terms than expected. They generally preferred intuitive media as alternatives to musical knowledge and universal interface during the prototype experience. These included UI elements such as emojis, control bars, onboarding phase, and progress bars depicted in Figure~\ref{fg:redesign}. There was a shared idea for the sample recommendation feature to facilitate intuitive designs within guided frames, associated by their experiences, such as EQ presets and design recommendation websites. P5 suggested that the current flow could be monotonous, so customizing samples like EQ presets could enhance the experience. P2 also mentioned the samples in certain design websites, which provides users with confirmation on the right direction.

\subsection{Explaining Sound in Collaboration}

The sound designers in our study predominantly had backgrounds in electrical engineering, but their work often requires collaboration with other departments, leading to an interdisciplinary environment. This dynamic involves explaining design concepts and allowing stakeholders to experience sonic outputs, resulting in variations of comprehension due to diverse backgrounds. This is not only unique within their company but also occurs with client firms. The sound designers explain and persuade clients to accept designed sounds through negotiations. P4 mentioned communicating with sound managers in client companies during the initial phase to discuss branding image and understanding common patterns within an interdisciplinary environment. P6 also highlighted negotiations with client companies through continuous listening and intuitive explanations. Similarly, both P5 and P6 noted using domain-specific formal language, such as variables and indexes, for documentation but employing informal descriptive language for immediate communication. These flexible linguistic conventions enable adept explainability in the interdisciplinary environment. The nature of work dynamics in designing sound entails the explanation for diverse collaboration while striving towards a common goal.

\subsection{Design Context of Sonic Interaction}

While sound design begins with a framed direction, it encounters various situations that require sophisticated optimization. P4 emphasized his company identities and design philosophies as an audio electronics firm, and there is even a potential for an additional layer of mutual consideration with client companies. P6 highlighted the sound color differences among client companies and addressing end user needs. Additionally, P1 underscored considering human and environmental characteristics as contextual variables, which was further mentioned by P4 and P5. P5 elaborated on the controller area network (CAN) as the systematic foundation for contextual possibilities. Similarly, during the focus group with sound designers, the necessity for contextual connections with sound and music variables was addressed. Lastly, P5 mentioned considering market trends along with end user needs. He provided an example of fuel efficiency in car weight reduction, where the subsequent question pertains to driving feeling, aligning with the emotional aspects mentioned by P4.

\section{Conclusion and Future Work}

Our study shows that although the sound design process is initially structured around relatively concrete foundations, ultimately, it must accommodate subjective and intuitive nuances. The final phase of sound design is characterized by qualitative factors and involves multiple layers of sound optimization. While audio engineering has its own domain knowledge, this can be adapted to the interdisciplinary dynamic of sound design. The sound designers we studied primarily communicate internally but also collaborate with other departments and external stakeholders. We found that a broader range of individuals than expected are involved in designing sound, indicating that it is not solely the domain of sound design specialists. These thematic frames overlap and influence one another, highlighting the mutual interactions taking place.

Our findings lead to the following recommendations for future development of an accessible tool that supports sound designers in the sonification with tonal cognition. Potential users for sound design software tool must include UX designers, not just sound design specialists, and could potentially extend to other design and engineering roles. The interface should be intuitive, but still involve domain-specific elements such as adding advanced options to the general task flow to suit both beginners and professionals. To ensure that the design direction aligns with the foundational settings, an onboarding phase is necessary, akin to the initial step of current sound design tasks. Similarly, sample suggestions would benefit designers as a guided framework. Lastly, musical knowledge could be described in engineering or common terminology for accessibility. Musical terms could be distilled into universally understood UI elements, such as major and minor tonal differences depicted by emojis and intuitive variable controllers like equalizers. The final outputs could be represented in a formal format for documentation, even though the interface language may be more intuitive and plain, resembling everyday communications observed in our study.

Future work could explore further details of the sound design software tool leveraging the suggested sonification method based on the findings by the current study. As part of this ongoing research, we will develop a high-fidelity prototype incorporating the discussed design directions. This development process will involve desk methods such as referencing similar software solutions, but the next phase will mainly entail usability evaluation of the refined prototype, possibly in a controlled experiment setting, and the data will be analyzed through both quantitative and qualitative analyses. The research and resultant sonic interaction design tool with tonal cognition will enable practitioners beyond exclusively sound specialists to create sound intuitively but logically with practical methods, infusing insights from the science of music and interaction design.

\bibliographystyle{ACM-Reference-Format}
\bibliography{reference}

\appendix

\section{APPENDIX: SUPPLEMENTARY MATERIALS}

\subsection{Semi-structured Interview Questions\label{appendix:question}}

These are the questions for individual and focus group interviews. The interviews comprehensively covered the categories below, but not every single question was necessarily asked of all participants or focus groups.

\subsubsection{Individual Interview\label{appendix:individual}}

\begin{itemize}
    \item How do they plan the design process?
    \begin{itemize}
        \item Can you outline the overall sound design process?
        \item Can you detail the ideation phase?
        \item Can you detail the prototyping phase? 
        \item What considerations should be reflected when planning the design process? 
    \end{itemize}
    \item Which design phases and tasks are significant?
    \begin{itemize}
        \item What are the particularly significant processes and tasks in ideation? 
        \item What are the particularly significant processes and tasks in prototyping? 
        \item Why do you think the processes and tasks are important? 
        \item Looking at the end-to-end process, where would you say are the critical decisions?
    \end{itemize}
    \item How do they imagine and decide design ideas?
    \begin{itemize}
        \item How do you come up with design ideas?
        \item How do you prioritize ideas and make final decisions? 
    \end{itemize}
     \item What are the significant design factors?
     \begin{itemize}
         \item What are the key sound design factors?
         \item Why are the factors important?
     \end{itemize}
    \item How do they connect ideas with design factors?
    \begin{itemize}
        \item Can you detail the factor mapping process?
        \item What are the differences and similarities between design ideas and factors?
        \item What are the difficulties in connecting design ideas and factors?
    \end{itemize}
     \item How do they prototype the sound?
     \begin{itemize}
         \item How do you fine-tune the sound output? 
         \item From which perspective are the produced sounds refined?
         \item What are the heuristics, are there rules of thumb you use to create sounds?
         \item How do you know the sound is suitable?
         \item Do musical foundations play a role in your process? 
         \item What software do you use? 
         \item Why do you use the software?
         \item What are the main features of the software you use? 
         \item Do they limit what you do, how so? 
         \item Do the digital tools you use provide music theoretical ideas to tune or modify the sounds?
     \end{itemize}
    \item How do they verbalize design ideas, factors, process, and sound outputs?
    \begin{itemize}
        \item What is the typical language used when brainstorming design ideas? 
        \item What is the common language or manner in which design factors are expressed? 
        \item What is the common language or manner in which sound outputs are described? 
        \item Are there instances where languages with distinct characteristics create challenges?
        \item Do you use different language with different collaborators?
        \item How did you acquire this language?
    \end{itemize}
    \item Overall opinions about the interdisciplinary work setting and current software
    \begin{itemize}
        \item What are the characteristics of sound design with people from diverse backgrounds?
        \item Do you work with others? Can you say what skills they contribute and how you work with them? 
        \item If they have a different background, how do you describe the sounds, what language do you use? 
        \item Are there distinctive aspects resulting from the diverse backgrounds of sound engineers? 
        \item Do you believe current software incorporates a wide range of sound design perspectives? 
        \item Do you think current software presents challenges depending on the designer's background? 
        \item What are the strong points and downsides of the software you use?
    \end{itemize}
\end{itemize}

\subsubsection{Focus Group\label{appendix:focusgroup}}

\begin{itemize}
    \item Examination of the low fidelity prototype in written and oral formats
    \begin{itemize}
        \item Are there any more design steps or tasks that were not considered? 
        \item Are there design phases or tasks that could be further connected back and forth? 
        \item What tasks can be further subdivided? 
        \item Are there any interface design points that would be good to consider? 
        \item Do you think it encompasses different design perspectives? 
        \item Do you think the applied language can be understood without advanced knowledge? 
        \item Do you think it connects intuitive design ideas with professional music knowledge? 
        \item Do you think it is an interface that people from various backgrounds can use? 
    \end{itemize}
    \item Overall discussion including broader design perspectives
    \begin{itemize}
        \item What do you think are the most essential design steps? 
        \item What characteristics of tasks do you think are key at the critical steps? 
        \item Why do you think connectivity between design phases is important? 
        \item What are the characteristics of words commonly used for design? 
        \item How do you develop your intuitive design ideas?
        \item How do you connect the design ideas with physical attributes?
        \item What do you consider for prototyping? 
        \item What is one feature you wish the software tool you use had? 
        \item How do you think practical, artistic or aesthetic, and physical perspectives are combined in design? 
        \item What are the pros and cons of having designers from different backgrounds working together? 
        \item Do you think current prototyping software reflects intuitive design ideas well? 
        \item What are the pros and cons of using current design and prototyping software in general? 
    \end{itemize}
\end{itemize}

\subsection{Low-Fidelity Prototype\label{append:figma}}

\begin{figure*}
  \centering
  \includegraphics[width=\textwidth]{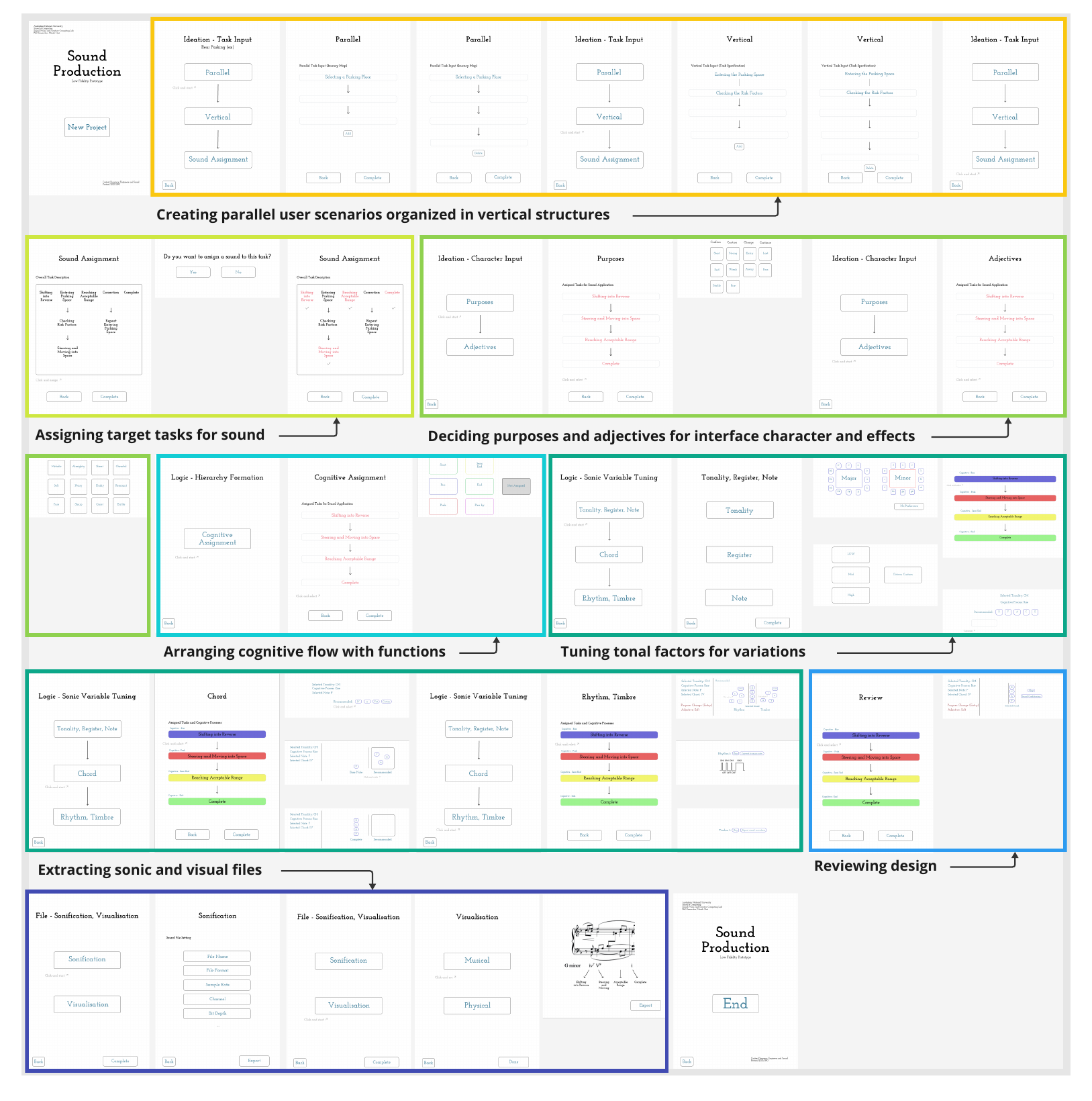}
  \caption{Low-fidelity prototype in Figma showing the overall prototype structure and the experience steps in focus group.}
  \label{fg:figma}
\end{figure*}

Figure~\ref{fg:figma} shows the low-fidelity prototype created in Figma. Different colors represent distinct design phases along with brief explanations. Some phases were simplified, such as only reviewing one sound rather than all the designed sounds.

\end{document}